\documentclass{article}
\usepackage{spconf,amsmath,graphicx}
\usepackage{multirow}
\usepackage{xcolor}
\usepackage{enumitem}
\usepackage{booktabs}
\usepackage{hyperref}

\title{Audio prompt tuning for universal sound separation}
\name{
Yuzhuo Liu$^{1,}\sthanks{Corrsponding author, yuzhuoliu1994@outlook.com}$,
Xubo Liu$^{2}$, 
Yan Zhao$^1$, 
Yuanyuan Wang$^3$, 
Rui Xia$^1$, 
Pingchuan Tain$^1$, 
Yuxuan Wang$^1$}
\address{ 
$^1$ Speech, Audio \& Music Intelligence (SAMI) Group, ByteDance Inc\\
$^2$ Centre for Vision, Speech and Signal Processing (CVSSP), University of Surrey\\
$^3$ Shenzhen International Graduate School, Tsinghua University \\}

\begin{document}
%
\maketitle
\begin{abstract}
 Universal sound separation (USS) is a task to separate arbitrary sounds from an audio mixture. Existing USS systems are capable of separating arbitrary sources, given a few examples of the target sources as queries. However, separating arbitrary sounds with a single system is challenging, and the robustness is not always guaranteed. In this work, we propose audio prompt tuning (APT), a simple yet effective approach to enhance existing USS systems. Specifically, APT improves the separation performance of specific sources through training a small number of prompt parameters with limited audio samples, while maintaining the generalization of the USS model by keeping its parameters frozen. We evaluate the proposed method on MUSDB18 and ESC-50 datasets. Compared with the baseline model, APT can improve the signal-to-distortion ratio performance by 0.67 dB and 2.06 dB using the full training set of two datasets. Moreover, APT with only 5 audio samples even outperforms the baseline systems utilizing full training data on the ESC-50 dataset, indicating the great potential of few-shot APT.

\end{abstract}
\begin{keywords}
Sound separation, audio prompt tuning, few-shot learning
\end{keywords}
\section{Introduction}
\label{sec:intro}

With the fast growth of deep learning in sound separation, universal sound separation (USS) has achieved remarkable progress in recent years~\cite{DBLP:conf/waspaa/WisdomJWEH21,DBLP:conf/icassp/KongWSCWP20, audiosep}. USS aims to extract arbitrary target sounds and can be used for applications such as audio editing and audio transcription. Most of the successful backbones focus on the capability of handling all kinds of general sounds~\cite{DBLP:conf/aaai/0021DZMBD22, DBLP:conf/interspeech/DelcroixVOKA21,DBLP:journals/corr/abs-2305-07447}. However, real-world applications may have higher performance demand on some frequently used categories. In this paper, we focus on the challenges of keeping the ability of general sound extraction while improving the performance on some seen and unseen sound classes with a small amount of class-specific data.

The framework of query-based USS is composed of two modules, condition embedding extraction, and target sound extraction~\cite{DBLP:conf/aaai/0021DZMBD22,delcroix2022soundbeam,chen2023iquery}. The condition embedding extraction module accepts enrollment audios and uses down-sampled utterance-level latent features from the last layers as conditional embedding. In the one-shot case, the conditional embedding equals the latent feature. In the few-shot case, the conditional embedding is calculated by averaging the features. The sound extraction module accepts the condition embedding as well as the mixture audio to extract the target sound, similar to the target speaker extraction~\cite{vzmolikova2019speakerbeam,DBLP:conf/interspeech/WangMWSWHSWJL19,yu23b_interspeech}. The condition extraction usually uses a pre-trained sound event detection (SED) network~\cite{DBLP:journals/taslp/KongCIWWP20,DBLP:conf/icassp/ChenDZMBD22} while the target sound extractor usually uses U-net-based models~\cite{DBLP:conf/aaai/0021DZMBD22,DBLP:conf/ismir/StollerED18}.

The two-stage framework can perform universal sound separation with enrollment audios. Nonetheless, the problem of the two-stage framework is two-fold. Firstly, the condition embeddings are trained for classification, leading to the mismatch problem, especially on the out-of-domain and unseen class samples. Secondly, the separation model is trained to collaborate with the embeddings from the SED networks. During inference, the extraction network needs to adapt to specific conditions which may produce imperfect performance since the model is forced to be capable of handling universal sources.

On the other hand, real-world applications require a higher extraction performance, especially on events that happen frequently or are remarkable in some situations. Meanwhile, the capability of dealing with universal sound categories needs to be kept as diverse sources need to be separated. Therefore, the important issue is how to improve the performance on specific event classes while keeping the original sound extraction ability unchanged. 

Inspired by the prompt tuning approaches in NLP~\cite{DBLP:conf/emnlp/LesterAC21,DBLP:conf/acl/LiL20,DBLP:conf/eccv/JiaTCCBHL22}, we propose audio prompt tuning (APT) for USS. With an assumption that a small amount of data is available for the specific event classes, the prompt tuning technique tunes the condition embedding for a pre-trained source separation model. The tuned prompts are adapted to the separation model, which solves the mismatch problem between condition embeddings and the separation model. Consequently, the tuned prompts achieve higher performance than fixed averaged embeddings on the specific event classes without changing the main separation model.

Experimental results demonstrate that the prompt tuning achieves better signal-to-distortion (SDR) compared to the original embeddings extracted from the classification model on both full training data and few-shot samples. Notably, these experiments are carried out on MUSDB18 and ESC-50 datasets that have different domains and contain unseen classes than the training data of the USS model. 

Our main contributions in this paper can be summarized in two aspects:

\begin{itemize}
[itemsep=0pt,topsep=0pt,parsep=0pt,leftmargin=10pt]
\item 
\textbf{APT} is designed for universal source separation. It tunes the condition embedding initialized by a classification model and gives a straightforward performance improvement compared with the classification embedding. The number of tuned parameters is less than 0.1\% of the parameters of the USS model.
\item 
\textbf{Data efficiency} is explored to indicate the performance of few-shot APT, which illustrates that APT achieves higher separation performance in all few-shot settings.
\end{itemize}

\section{Method}
\label{sec:method}

\begin{figure}[th]
\centering{\includegraphics[width=0.55\linewidth]{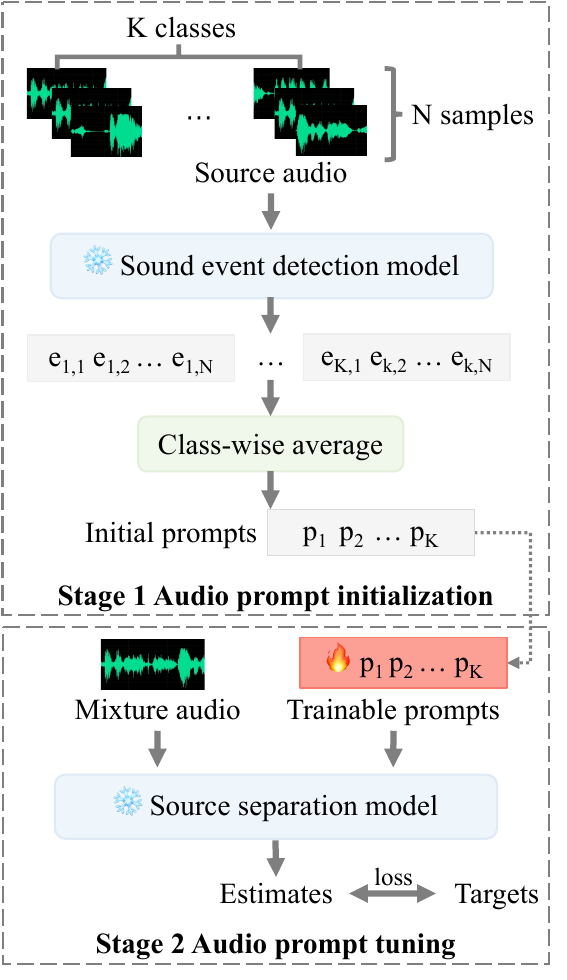}}
\caption{Overview of two-stage APT framework for USS. Initial prompts are generated by the SED model in Stage 1, and only the parameters of prompts are tuned in Stage 2. The parameters of SED and USS models are frozen.}
\label{fig:pipeline}
\end{figure}

Figure \ref{fig:pipeline} plots the pipeline of the proposed APT. The whole procedure consists of two stages: audio prompt initialization and audio prompt tuning. This section gives a detailed description of the two-stage pipeline and includes some discussion of the prompt tuning in the USS task.

\subsection{Stage 1: Audio prompt initialization}
The first stage generates the initial embedding. For the K-class N-shot dataset, the $i$th sample from class $k$ is denoted as $x_{k,i}$. Samples are first fed into a pre-trained SED model to generate the utterance-level embeddings,
\begin{equation}
    e_{k,i} = \text{SED}(x_{k,i}),
\end{equation}
where $e_{k,i}$ is the calculated embedding of sample $i$ in class $k$; $\text{SED}$ is a HTS-AT model pre-trained on the AudioSet \cite{DBLP:conf/icassp/ChenDZMBD22}. Average pooling is adapted to generate the initial class-wise prompts,
\begin{equation}
    p_k = \sum_i e_{k,i} / N.
\end{equation}
In zero-shot query-based USS, the initialized embedding is directly sent to the separation model as the query embedding. However, our proposed prompt tuning further processes the initialized embedding in the second stage.

\subsection{Stage 2: Audio prompt tuning}

In the second stage, we first use the K-class N-shot dataset to simulate a mixture dataset, where the mixture audio is obtained by:
\begin{equation}
    y_{k,i} = x_{k,i} + x_{l,j},
\end{equation}
where $k\neq l$. The training target of prompt tuning is to recover $x_{k,i}$ from the mixture observation $y_{k,i}$. Specifically, the USS model takes the mixture and the prompt as inputs to estimate $x_{k,i}$, namely,
\begin{equation}
    \hat{x}_{k,i} = \text{USS}(y_{k,i}|\Theta_{USS},p_k),
\end{equation}
where $\Theta_{USS}$ denotes the parameters of the separation model; $p_k$ is the tuning prompt initialized from the first stage. We optimize the model by utilizing waveform-based L1 loss between the estimated $\hat{x}_{k,i}$ and it corresponding clean reference $x_{k,i}$.
In practice, we only train the input prompt $p_k$ and freeze $\Theta_\text{USS}$. In such a way, the original ability of zero-shot query-based separation remains while the performance of the important classes is improved.

\subsection{Discussion and comparison}

Prompt tuning is a simple yet effective and general approach to enhance separation performance. It differs from zero-shot query-based \cite{DBLP:conf/aaai/0021DZMBD22} and one-hot-based separation \cite{DBLP:conf/ismir/Meseguer-Brocal19}. Zero-shot query-based separation replies on initializing the input query embedding with samples from the target class. However, such initialization might yield suboptimal queries due to out-of-domain data distribution and the mismatch between classification and separation tasks. The proposed prompt tuning directly addresses these issues by fine-tuning the prompts. In contrast to one-hot-based separation, the proposed APT benefits from a superior initialization instead of training embeddings from scratch. Meanwhile, it retains the generalization capabilities of query-based USS.

It is worth noting that the proposed prompt tuning is not tied to specific model architectures, making it adaptable to various query-based separation models.

\section{Expermental setup}
\label{sec:Exp}

\subsection{Dataset and evaluation}

Experiments are carried out on MUSDB18~\cite{rafii2017musdb18} and ESC-50~\cite{piczak2015dataset} datasets. The original USS model was trained using AudioSet. Thus, MUSDB18 and ESC-50 are used to verify the feasibility of the proposed prompt tuning to different domain data under limited resource scenarios.

 \textbf{MUSDB18}  dataset focuses on the music separation task, which consists of a training set (100 samples) and a test set (50 samples). Every sample includes four source tracks (vocal, drum, bass, other) and their mixture. We use the source tracks in the training set to initialize the latent embedding in the first stage, as shown in Figure \ref{fig:pipeline}. All training samples are adopted and are randomly cropped into two-second clips during training~\cite{DBLP:conf/aaai/0021DZMBD22}.

\textbf{ESC-50} dataset consists of 50 categories of sound events from nature, water, human non-speech, domestic and urban soundscapes. Each event class contains 40 5-second samples. These audio samples are officially divided into 5 folds. The mixtures are generated by mixing each sample with another one from a different event class. We train the prompts with the mixtures simulated from Fold 1 to 3. The sources in Fold 4 to 5 are used to generate the test mixtures.

The signal-to-distortion ratio (SDR) is utilized to evaluate our method. We use median SDR to follow the settings in \cite{DBLP:conf/aaai/0021DZMBD22}. The average score is the median over all mixture audio clips.

\subsection{Baseline and settings}

We choose the 2048-d ST-SED-SEP model~\cite{DBLP:conf/aaai/0021DZMBD22} as the baseline and backbone model. The 2048-d ST-SED-SEP model uses the averaged embedding as an input condition to train and infer the separation model directly. For a fair comparison, we adopt the same SED model and USS model as the baseline model, and the parameters of both models are not involved in training. It is worth noting that the USS model has 180M parameters and the trainable parameter for prompt tuning is only 8.2K and 102.4K for 4-class MUSDB18 and 50-class ESC-50.

All input audios are resampled to 32kHz. Then, these audios are transformed into Mel features with 1024 window size, 320 hop size, and 64 bins. The dimension of the prompt embedding is 2048.

The learning rate and batch size are set to $3\times 10^{-4}$ and 4, respectively. We train 500 and 100 epochs using the simulated training data from MUSDB18 and ESC-50, respectively.

\section{Results and discussion}
\label{sec:results}

\subsection{Separation results with the full training set}
\begin{figure*}[t]
\centering{\includegraphics[width=0.95\linewidth]{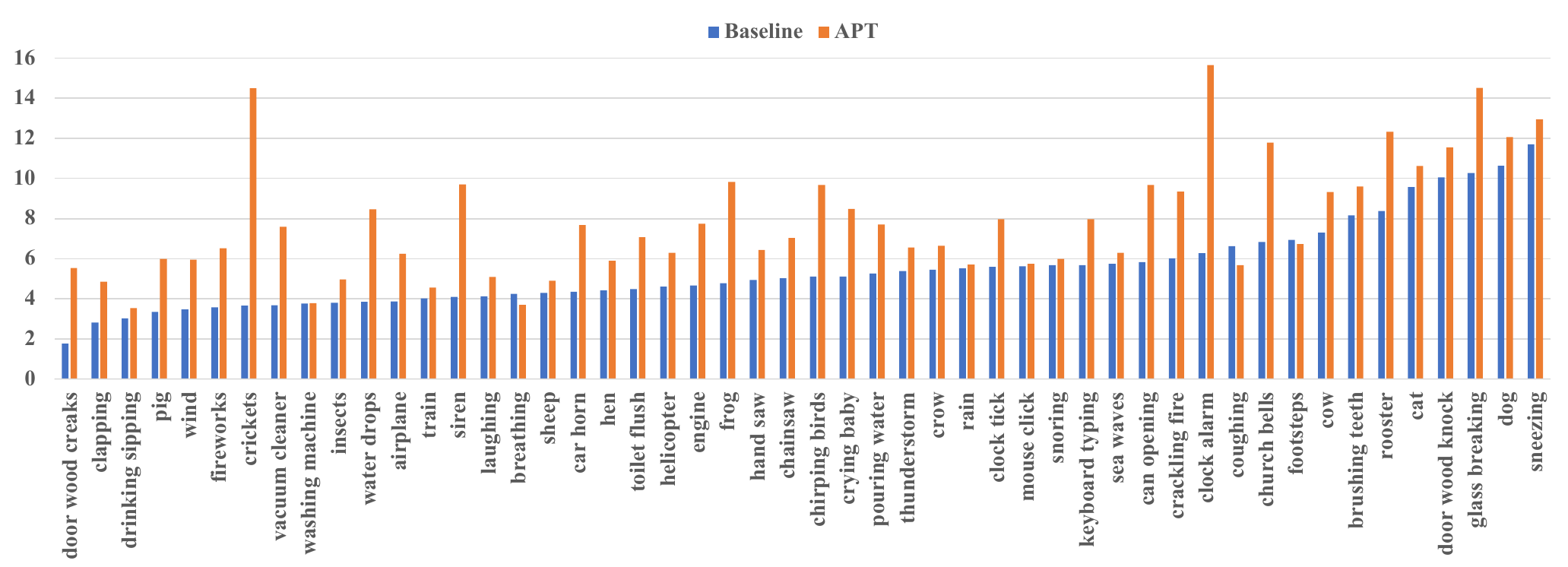}}
\caption{The SDRs of the APT and baseline model of all 50 events on the ESC-50 dataset. Models are trained with a full training set. \textbf{APT achieves higher SDRs than the baseline in 47 classes out of 50.} }
\label{fig:50classes}
\end{figure*}

The comparison of the SDR scores of the proposed APT and baseline model with the full training set is carried out on both MUSDB18 and ESC-50 datasets. Table~\ref{tab:full_set} shows the detailed SDRs of all instrument classes on MUSDB18 and the average SDR scores of two datasets. Figure~\ref{fig:50classes} illustrates the performance of all 50 categories on ESC-50. 

\begin{table}[ht]
\caption{Comparison of the SDR scores of the proposed APT, baseline model, joint fine-tuning of the prompts and the USS model with full training data of the MUSDB18 and ESC-50 datasets.}
\vspace{0.5em}
\label{tab:full_set}
\centering
\begin{tabular}{ccccc}
\toprule
\textbf{Dataset}               &\textbf{Class}           & \textbf{Baseline} & \textbf{APT}  &\textbf{Joint-Tune} \\ \midrule
\multirow{5}{*}{MUSDB18} & Vocal     & 6.15         & 6.31   & 6.42       \\
                         & Drum      & 5.44         & 6.36      & 6.41   \\
                         & Bass      & 3.80         & 4.92     & 5.11     \\
                         & Other     & 3.05         & 3.12    &  3.93    \\
                         & Average & 4.31           & 4.98    &  5.02      \\ \midrule
ESC-50                   & Average & 6.44           & 8.50      &9.32    \\ \bottomrule
\end{tabular}
\end{table} 

Table~\ref{tab:full_set} and Figure~\ref{fig:50classes} demonstrate that the APT significantly outperforms the baseline method in the scenario where the full training set is employed. Firstly, the average SDR scores of APT surpass the baseline model by 0.67 dB and 2.06 dB on two datasets, respectively. Secondly, APT exceeds the 2048-d ST-SED-SEP on all 4 instruments of MUSDB18. Out of 50 classes on ESC-50, the separation performance of 47 events has been improved, and the improvement of 38 events has exceeded 1 dB. We observe that the unseen classes (Drinking sipping, mouse click, can opening, washing machine, door wood creaks) in AudioSet are all included in the 47 improved sources, and two of them (can opening, door wood creaks) achieve 2-3 dB improvements.

We fine-tune the prompts and the USS model jointly to explore the upper-bound performance. Results in Table~\ref{tab:full_set} illustrate that the APT loses only 0.04 dB and 0.82 dB on two datasets, respectively. Meanwhile, APT can keep the generalization of the USS.

\subsection{Few-shot verification}
 To evaluate the performance in few-shot scenarios, N samples of each class are selected randomly to form new N-shot training sets. In this study, we investigate the cases where N=1, 5, and 10. The 10-shot training set includes samples in the 5-shot set, and the 5-shot set includes the sample in the 1-shot set. We use these new few-shot sets to initialize the classification embedding and train prompts. Few-shot and full-data experiments share the same test set.

Table~\ref{tab:few-shot} illustrates that APT achieves higher average SDRs than the scores of the baseline model under all few-shot settings. Both the performance of the baseline and APT are improved as training samples increase, while APT always achieves higher SDRs. Moreover, the APT trained with only 5 samples is observed to surpass embeddings initialized with all 24 samples. Detailed few-shot results of 50 events are available online~\footnote{\url{https://github.com/redrabbit94/APT-USS/blob/main/Results-ESC50.csv}}. In 1-shot, 5-shot, 10-shot experiments, 30, 36, 42 events out of 50 obtain better performance. This phenomenon suggests the benefit of prompt tuning and indicates that more training samples bring greater improvements.

\begin{table}[t]
\caption{The SDRs of the APT and baseline model with different few-shot settings on the ESC-50 dataset.}
\vspace{0.5em}
\label{tab:few-shot}
\centering
\begin{tabular}{lcccc}
\toprule
        & \textbf{1-shot} & \textbf{5-shot} & \textbf{10-shot} & \textbf{Full-data}  \\ \midrule
\textbf{Baseline}    & 4.09   & 5.59    & 6.10   &6.44        \\
\textbf{APT}               & 4.57   & 6.68   & 7.59    & 8.50       \\ \bottomrule           
\end{tabular}
\end{table}

\begin{figure}[t]
\centering{\includegraphics[width=80mm]{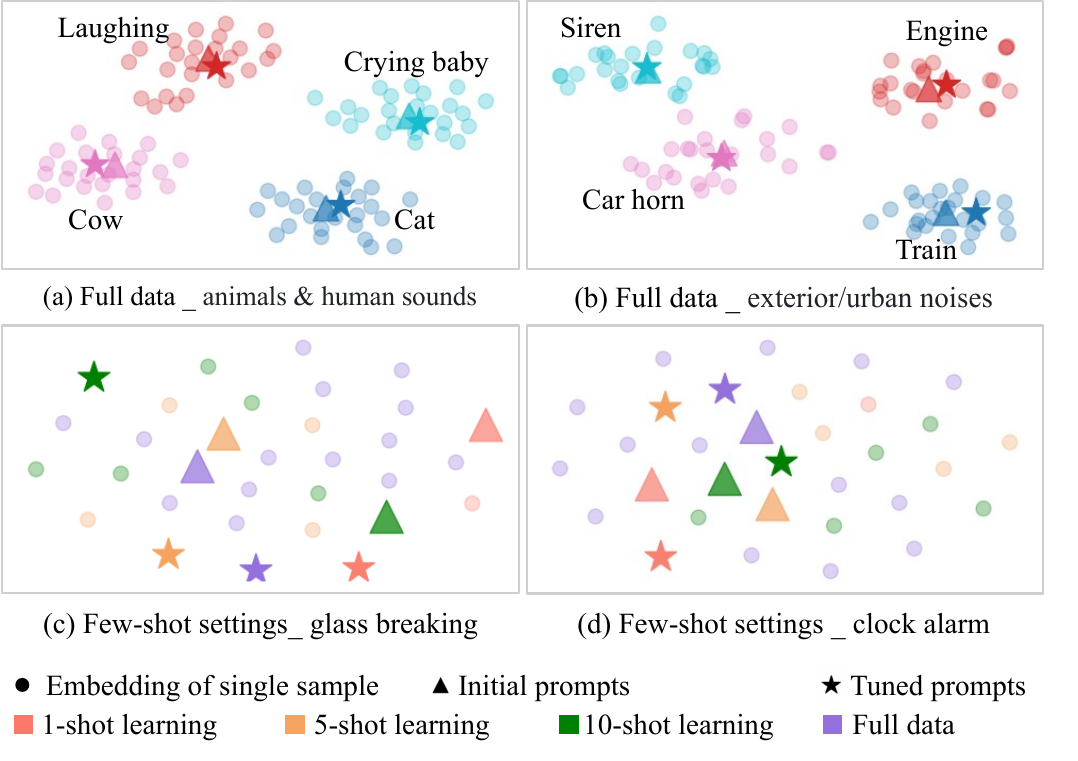}}
\caption{T-SNE visualizations of sample embeddings, initial prompts and tuned prompts for different classes (using different colors in (a)-(b) ) and few-shot conditions (using different colors in (c)-(d) ). Compared with the initial prompts used in the baseline model, tuned prompts are shifted. }
\label{fig:scatters}
\end{figure}

\subsection{Visualization\quad}
Figure \ref{fig:scatters} presents the visualization of different classes and the shift of prompts in the few-shot learning. Figure \ref{fig:scatters}(a)-(b) plots the sounds from human beings and animals, and the sounds from machines. Each shows clear boundaries among different classes, indicating that embeddings from the classification model exhibit intra-class consistency and inter-class discrimination. The prompts after tuning are shifted compared with the initial prompts but still located within the class boundaries, suggesting that directly tuning the prompts will improve the performance onto local optimal.

Figure \ref{fig:scatters}(c)-(d) shows the shift of initial and tuned prompts under different few-shot settings. Most shifts between the initial and tuned prompts exhibit obvious directions, implying the optimizing gradient for prompt tuning.

\section{Conclusion}
\label{sec:conclusion}
This paper proposes a simple APT method for USS. The APT only trains a few parameters with limited data that can be easily applied in various scenarios. Our experiments illustrate that the APT method can effectively improve separation performance while keeping the generalization of the backbone USS model. More prompt tuning techniques for USS and speech extraction will be studied and compared in the future.

\bibliographystyle{IEEEtran}
\bibliography{strings,refs}

\begin{thebibliography}{10}
\providecommand{\url}[1]{#1}
\csname url@samestyle\endcsname
\providecommand{\newblock}{\relax}
\providecommand{\bibinfo}[2]{#2}
\providecommand{\BIBentrySTDinterwordspacing}{\spaceskip=0pt\relax}
\providecommand{\BIBentryALTinterwordstretchfactor}{4}
\providecommand{\BIBentryALTinterwordspacing}{\spaceskip=\fontdimen2\font plus
\BIBentryALTinterwordstretchfactor\fontdimen3\font minus
  \fontdimen4\font\relax}
\providecommand{\BIBforeignlanguage}[2]{{%
\expandafter\ifx\csname l@#1\endcsname\relax
\typeout{** WARNING: IEEEtran.bst: No hyphenation pattern has been}%
\typeout{** loaded for the language `#1'. Using the pattern for}%
\typeout{** the default language instead.}%
\else
\language=\csname l@#1\endcsname
\fi
#2}}
\providecommand{\BIBdecl}{\relax}
\BIBdecl

\bibitem{DBLP:conf/waspaa/WisdomJWEH21}
S.~Wisdom, A.~Jansen, R.~J. Weiss, H.~Erdogan, and J.~R. Hershey, ``Sparse,
  efficient, and semantic mixture invariant training: Taming in-the-wild
  unsupervised sound separation,'' in \emph{{IEEE} Workshop on Applications of
  Signal Processing to Audio and Acoustics}, 2021, pp. 51--55.

\bibitem{DBLP:conf/icassp/KongWSCWP20}
Q.~Kong, Y.~Wang, X.~Song, Y.~Cao, W.~Wang, and M.~D. Plumbley, ``Source
  separation with weakly labelled data: an approach to computational auditory
  scene analysis,'' in \emph{{ICASSP}}.\hskip 1em plus 0.5em minus 0.4em\relax
  {IEEE}, 2020, pp. 101--105.

\bibitem{audiosep}
X.~Liu, Q.~Kong, Y.~Zhao, H.~Liu, Y.~Yuan, Y.~Liu, R.~Xia, Y.~Wang, M.~D.
  Plumbley, and W.~Wang, ``Separate anything you describe,'' \emph{arXiv
  preprint arXiv:2308.05037}, 2023.

\bibitem{DBLP:conf/aaai/0021DZMBD22}
K.~Chen, X.~Du, B.~Zhu, Z.~Ma, T.~Berg-Kirkpatrick, and S.~Dubnov, ``Zero-shot
  audio source separation through query-based learning from weakly-labeled
  data,'' in \emph{Proceedings of the AAAI Conference on Artificial
  Intelligence}, vol.~36, no.~4, 2022, pp. 4441--4449.

\bibitem{DBLP:conf/interspeech/DelcroixVOKA21}
M.~Delcroix, J.~B. V{\'{a}}zquez, T.~Ochiai, K.~Kinoshita, and S.~Araki,
  ``Few-shot learning of new sound classes for target sound extraction,'' in
  \emph{Interspeech, Brno, Czechia, 30 August - 3 September 2021}.\hskip 1em
  plus 0.5em minus 0.4em\relax {ISCA}, 2021, pp. 3500--3504.

\bibitem{DBLP:journals/corr/abs-2305-07447}
Q.~Kong, K.~Chen, H.~Liu, X.~Du, T.~Berg{-}Kirkpatrick, S.~Dubnov, and M.~D.
  Plumbley, ``Universal source separation with weakly labelled data,''
  \emph{CoRR}, vol. abs/2305.07447, 2023.

\bibitem{delcroix2022soundbeam}
M.~Delcroix, J.~B. V{\'a}zquez, T.~Ochiai, K.~Kinoshita, Y.~Ohishi, and
  S.~Araki, ``Soundbeam: Target sound extraction conditioned on sound-class
  labels and enrollment clues for increased performance and continuous
  learning,'' \emph{IEEE/ACM Transactions on Audio, Speech, and Language
  Processing}, vol.~31, pp. 121--136, 2022.

\bibitem{chen2023iquery}
J.~Chen, R.~Zhang, D.~Lian, J.~Yang, Z.~Zeng, and J.~Shi, ``iquery: Instruments
  as queries for audio-visual sound separation,'' in \emph{Proceedings of the
  IEEE/CVF Conference on Computer Vision and Pattern Recognition}, 2023, pp.
  14\,675--14\,686.

\bibitem{vzmolikova2019speakerbeam}
K.~{\v{Z}}mol{\'\i}kov{\'a}, M.~Delcroix, K.~Kinoshita, T.~Ochiai, T.~Nakatani,
  L.~Burget, and J.~{\v{C}}ernock{\`y}, ``Speakerbeam: Speaker aware neural
  network for target speaker extraction in speech mixtures,'' \emph{IEEE
  Journal of Selected Topics in Signal Processing}, pp. 800--814, 2019.

\bibitem{DBLP:conf/interspeech/WangMWSWHSWJL19}
Q.~Wang, H.~Muckenhirn, K.~W. Wilson, P.~Sridhar, Z.~Wu, J.~R. Hershey, R.~A.
  Saurous, R.~J. Weiss, Y.~Jia, and I.~Lopez{-}Moreno, ``Voicefilter: Targeted
  voice separation by speaker-conditioned spectrogram masking,'' in
  \emph{Interspeech}, 2019, pp. 2728--2732.

\bibitem{yu23b_interspeech}
J.~Yu, H.~Chen, Y.~Luo, R.~Gu, and C.~Weng, ``{High Fidelity Speech Enhancement
  with Band-split RNN},'' in \emph{Proc. INTERSPEECH 2023}, 2023, pp.
  2483--2487.

\bibitem{DBLP:journals/taslp/KongCIWWP20}
Q.~Kong, Y.~Cao, T.~Iqbal, Y.~Wang, W.~Wang, and M.~D. Plumbley, ``Panns:
  Large-scale pretrained audio neural networks for audio pattern recognition,''
  \emph{{IEEE} {ACM} Trans. Audio Speech Lang. Process.}, vol.~28, pp.
  2880--2894, 2020.

\bibitem{DBLP:conf/icassp/ChenDZMBD22}
K.~Chen, X.~Du, B.~Zhu, Z.~Ma, T.~Berg{-}Kirkpatrick, and S.~Dubnov,
  ``{HTS-AT:} {A} hierarchical token-semantic audio transformer for sound
  classification and detection,'' in \emph{{ICASSP}}.\hskip 1em plus 0.5em
  minus 0.4em\relax {IEEE}, 2022, pp. 646--650.

\bibitem{DBLP:conf/ismir/StollerED18}
D.~Stoller, S.~Ewert, and S.~Dixon, ``Wave-u-net: {A} multi-scale neural
  network for end-to-end audio source separation,'' in \emph{Proceedings of the
  19th International Society for Music Information Retrieval Conference}, 2018,
  pp. 334--340.

\bibitem{DBLP:conf/emnlp/LesterAC21}
B.~Lester, R.~Al{-}Rfou, and N.~Constant, ``The power of scale for
  parameter-efficient prompt tuning,'' in \emph{Proceedings of the 2021
  Conference on Empirical Methods in Natural Language Processing, Virtual Event
  / Punta Cana, Dominican Republic}.\hskip 1em plus 0.5em minus 0.4em\relax
  Association for Computational Linguistics, 2021, pp. 3045--3059.

\bibitem{DBLP:conf/acl/LiL20}
X.~L. Li and P.~Liang, ``Prefix-tuning: Optimizing continuous prompts for
  generation,'' in \emph{Proceedings of the 59th Annual Meeting of the
  Association for Computational Linguistics and the 11th International Joint
  Conference on Natural Language Processing (Volume 1: Long Papers)}, 2021, pp.
  4582--4597.

\bibitem{DBLP:conf/eccv/JiaTCCBHL22}
M.~Jia, L.~Tang, B.~Chen, C.~Cardie, S.~J. Belongie, B.~Hariharan, and S.~Lim,
  ``Visual prompt tuning,'' in \emph{Computer Vision - {ECCV} 2022}, vol.
  13693.\hskip 1em plus 0.5em minus 0.4em\relax Springer, 2022, pp. 709--727.

\bibitem{DBLP:conf/ismir/Meseguer-Brocal19}
G.~Meseguer{-}Brocal and G.~Peeters, ``Conditioned-u-net: Introducing a control
  mechanism in the u-net for multiple source separations,'' in
  \emph{Proceedings of the 20th International Society for Music Information
  Retrieval Conference, Delft, The Netherlands, November 4-8}, 2019, pp.
  159--165.

\bibitem{rafii2017musdb18}
Z.~Rafii, A.~Liutkus, F.-R. St{\"o}ter, S.~I. Mimilakis, and R.~Bittner,
  ``Musdb18-a corpus for music separation,'' 2017.

\bibitem{piczak2015dataset}
K.~J. Piczak, ``{ESC}: {Dataset} for {Environmental Sound Classification},'' in
  \emph{Proceedings of the 23rd {Annual ACM Conference} on {Multimedia}}, pp.
  1015--1018.

\end{thebibliography}
\end{document}